\documentstyle[preprint,prb,eqsecnum,aps]{revtex}

\begin{document}
\tightenlines
\preprint{To appear in J. Chem. Phys.}


\title{Spontaneous curvature-induced dynamical instability of
Kirchhoff filaments: Application to DNA kink deformations}

\author{Zhou Haijun$^{1,}$\cite{zhou} and 
Ou-Yang Zhong-can$^{1,2}$}

\address{ 
$^1$Institute of Theoretical Physics, Academia Sinica, 
P.O. Box 2735 Beijing 100080, China\\
$^2$Center for Advanced Study, Tsinghua University, Beijing 100084, China
}
\date{September 14, 1998}
\maketitle


\begin{abstract}
The Kirchhoff elastic theory of thin filaments with spontaneous
curvature is employed in the understanding of the onset of  the kink transitions
 observed in short DNA
rings. Dynamical analysis shows that when its actual curvature
 is less than some threshold
 value determined by the spontaneous curvature, a circular DNA 
  will begin to buckle
to other shapes. The observable and
the dominant deformation modes are  also determined by 
 dynamical instability analysis, and
  the different effects of Zn$^{2+}$ and
Mg$^{2+}$ ions on DNA configurational properties are qualitatively discussed. 
\end{abstract}

\newpage


This paper is motivated mainly by a recent  experiment on short DNA 
rings. \cite{RA,RB}  In this experiment, Han {\it et al.} synthesized a 
 kind of DNA sequences 
 known to have intrinsic bending tendency (in its undistorted
 state  this kind of sequences  will form an axial bend
of at least $30^{\circ}$ per helical turn. \cite{RC})  They
used these sequences to investigate whether  DNA can
actually be kinked (forming large bendings over only few base pairs) and,
if kinks really turn up, what are the necessary conditions to produce them.
\cite{RA,RB} Kink deformation has been theoretically proposed to be a 
very important
mechanism for wrapping DNA around the nucleosome particles ever since the
seventies, \cite{RD,RE,RF}  and  a recent x-ray crystalline structure of
 nucleosome core particle
also demonstrated  that DNA is not uniformly 
bent but has maximal and minimal
curvatures at different positions.\cite{RG}  However, there had been no direct evidence
for the existence of kinks in DNA before the experiment
 of Han {\it et al.} \cite{RA,RB} which clearly demonstrate this. To their
 great surprise,  they found that kink formation is 
 closely related to
 total chain length and solvent ionic conditions.
For DNA circles with $168$ base pairs (bps) or $16$ helical turns,
 the chain is often kinked and form polygonal shapes if the solution has an 
 appropriate  Zn$^{2+}$ or Zn$^{2+}$/Mg$^{2+}$ concentrations. 
However, kink transition does not
take place in  Mg$^{2+}$ solution alone  or in Zn$^{2+}$/Mg$^{2+}$ solution
with high concentration of Mg$^{2+}$ ions.
 For DNA circles of $126$-bp or $12$ helical turns,
 no kink will ever occur, no matter what kind of 
 solvent ionic conditions. \cite{RA,RB}
Trying to interpret their novel observations, 
Han {\it et al.} suggested that  axial
stresses  exist in $168$-bp DNAs and they are the main   reason for
 the observed kink deformations.


In this article we attempt to 
check, from a theoretical point of view, 
 the validity of  this insight and try to understand the
radically different roles played by Mg$^{2+}$ and Zn$^{2+}$ ions in 
DNA kink transition.
 The spontaneous curvature   \cite{RC}  of the DNA material used
is anticipated to  play a significant role and we 
model the DNA chain as
a thin Kirchhoff elastic filament \cite{RH} with spontaneous curvatures,
 i.e., adjacent
cross-sections of the filament are tilted 
a certain angle with respect to each other (in a
mean-field sense).
The filament will  curve to  a planar ring of curvature 
$\kappa_{0}$ in its natural
undistorted state, as what is observed in experiments. \cite{RC} 
Our study shows that a nonzero spontaneous curvature is required
for the onset of the kink transitions observed in short DNA rings,  in consistency
with the insight of Han {\it et al.}
 
We define a moving orthonormal
coordinate system $\{{\bf d}_1 (s, t),{\bf d}_2(s, t), {\bf d}_3(s, t)\}$ 
along the filament axis line, with
${\bf d}_3$ being the  tangential unit vector at arc length $s$.
 In general the filament configuration will change with
time $t$, \cite{RI,RJ}  so the direction vectors 
are also functions of $t$. The 
choice of the other two unit vectors 
${\bf d}_1$ and ${\bf d}_2$ is arbitrary as long as 
${\bf d}_1 \times {\bf d}_2 ={\bf d}_3$; in our case we  define ${\bf d}_1$
to point to the tilt direction of the filament cross-section
at each arc length point $s$, for convenience. 
  Then the  free energy functional of such a filament
is equivalent to the following form
\begin{equation}
H={E I \over 2} \int
\left((\Omega_1-\kappa_{0})^2+ \Omega_2^2+\Gamma (\Omega_3-\omega_0)^2 \right) d s
\end{equation}
where $E$ is the Young's modulus of the filament and $I$ is the momentum of
inertia of the filament cross-section  (assumed to be circular),\cite{RK}
 $\Gamma$ is
a dimensionless parameter ranging between $2/3$ and $1$. \cite{RJ,RK} 
${\bf \Omega}=\sum_i
\Omega_i {\bf d}_i$ is called the twist vector,
 defined by the following equation 
$$
{\bf d}_i^{\prime}={\bf \Omega} \times {\bf d}_i,\;\;\;\;\;\;(i=1, 2, 3)
$$
here and after $(\;)^{\prime}$ means $s$-derivative. We have chosen such a 
continuous and homogeneous
 model partly because of the experimental fact that kinking
locations are not (or only very weakly)  correlated with base pair 
types. \cite{RB}

In proposing Eq.\ (1) we have also
take into account via the spontaneous twist rate $\omega_0$ the fact that
ordinary DNA can also have a linking number deficiency of $5\%$.\cite{R12} 
For DNA circles with $168$ bps ($16$ turns), it may well have $0.8$ turns of
initial twist in the DNA circle; for those with $126$ bps ($12$ turns), it
may well have $0.6$ turns of initial twist. However, this (possible) spontaneous twisting
tendency cannot explain the above mentioned kink deformations of short DNA. The
reason is the following: 
For an elastic filament formed by two chains
 interwinding around each other,  such
as DNA and actin,
a  well known model\cite{R13,R14,R15} concerning the effect of the fixed linking number
 shows that
 when the total twisting number variance $\Delta Tw$  exceeds some threshold
value $Tw_c =\sqrt{n^2 -1} A/C\;\;(n\geq 2)$, a
planar circle will buckle to the $n$-th deformation mode (this relation is also rederived in Appendix B), where
$A$ and $C$ are respectively the bending and twisting persistence length
(for DNA  $A \simeq 50$ nm and $C\simeq 75$ nm.) 
If the kink deformations are indeed induced by this topological effect, then for the typical 
square ($n =4$) polygonal kinked shape observed in the references\cite{RA,RB} to
appear  the total twisting number of the
$168$-bp DNA ring are required to deviate from its equilibrium value
  up to  $Tw_c=3$
turns. \cite{R13,R14,R15} This value is almost four times the actual value of $0.8$ turns. Thus
it seems that the spontaneous twist will only play a neglectable role in the kink deformations of short DNA,
 and we feel it might be appropriate to focus on 
the macroscopic bending tendency of the filament by setting $\omega_0 =0$ in Eq.\ 1. 
The results derived based on this approximation also
confirm this to be reasonable, as can be seen later. 
To be complete, we have also listed out the general result of model (1) in the case  of $\omega_0 \neq 0$ 
at the end of this paper (Appendix B), it reduces to the well known result in the references\cite{R13,R14,R15} for the
limiting case of DNA with intrinsic twist but no intrinsic bend ($\kappa_{0}=0$).


Recently Goriely, Tabor\cite{RJ} (and others)
 have suggested a powerful way to investigate 
 on the dynamical properties of Kirchhoff thin filaments.  We will use their
 procedure in studying the stabilities of  model (1). 
 Goriely and Tabor have worked out the stationary shape equations (SSEs)
 and the dynamical variational equations (DVEs) \cite{RJ}
 for the simplest case of a filament 
 without any spontaneous curvature. Here, for our purpose, we first give
the  SSEs and DVEs for a general Kirchhoff filament. (Since reference \cite{RJ} 
has  demonstrated a way to get the DVEs we
 will not waste space in writing down the
detailed calculations, but only notice here 
that there are some typographical errors in this reference.)
After the general SSEs and DVEs are obtained, we then turn back to the
 special case of Eq.\ (1). 

The configurational free energy functional for a general Kirchhoff filament  is  
 \cite{RH,RI,RJ,RK}
$$
H={E I\over 2}\int ((\Omega_1- K_1^e)^2+(\Omega_2-K_2^e)^2+\Gamma 
(\Omega_3-K_3^e)^2)ds,
$$
and the internal torque ${\bf M}$ is related to the deformation via the
following constitutive equation
$$
{\bf M}=E I (\Omega_1-K_1^e){\bf d}_1+E I (\Omega_2-K_2^e){\bf d}_2+
E I \Gamma (\Omega_3-K_3^e){\bf d}_3,
$$
here $K_i^e$ $(i=1, 2, 3)$ is the spontaneous curvature along the ${\bf d}_i$ 
direction.  There are also internal forces ${\bf F}$ along the filament.
\cite{RJ,RK}  We can perform a
standard scaling as listed in Eq.\ (19) of reference \cite{RJ}  to transform
 all the concerned quantities such as 
${\bf M}$ and  ${\bf F}$ into dimensionless forms. After this operation,
we perform a basis perturbation operation suggested by 
Goriely and Tabor \cite{RJ} to the stationary configuration ${\bf d}_i^{(0)}$
and get that
\begin{eqnarray}
{\bf d}_i&=&{\bf d}_i^{(0)}+\epsilon \;{\bf \alpha}\times {\bf d}_i^{(0)},\\
{\bf F}&=&f_i^{(0)} {\bf d}_i^{(0)}+\epsilon \; (f_i^{(1)} {\bf d}_i^{(0)}
+f_i^{(1)} {\bf \alpha}\times {\bf d}_i^{(0)}), \\
{\bf M}&=&(\Omega_i^{(0)}-K_i^e){\bf d}_i^{(0)}+\epsilon 
\;(\alpha_i^{\prime}{\bf d}_i^{(0)}-
\epsilon_{ijk}\alpha_i\Omega_j^{(0)}\Omega_k^{(0)}),
\end{eqnarray}
where $(\;)^{(0)}$ means  value corresponding to the stationary state 
and $(\;)^{(1)}$ its
first-order correction, $\epsilon$ is a small quantity. Insert
 Eqs.\ (2-4) into the Kirchhoff equations developed in  references \cite{RI,RJ} and 
after a lengthy but straightforward calculation we can obtain the SSEs to be:
\begin{eqnarray}
&&({\bf F}^{(0)})^{\prime\prime}=(f_1^{(0)}{\bf d}_1^{(0)}+f_2^{(0)}{\bf
 d}_2^{(0)}+f_3^{(0)}{\bf d}_3^{(0)})^{\prime\prime}=0,\\
&&(\Omega_1^{(0)}-K_1^e)^{\prime}-(\Omega_2^{(0)}-K_2^e) \Omega_3^{(0)}+
\Gamma (\Omega_3^{(0)}-K_3^e) \Omega_2^{(0)}=f_2^{(0)},\\
&&(\Omega_2^{(0)}-K_2^e)^{\prime}-\Gamma (\Omega_3^{(0)}-K_3^e)\Omega_1^{(0)}+
(\Omega_1^{(0)}-K_1^e)\Omega_3^{(0)}=-f_1^{(0)},\\
&&\Gamma (\Omega_3^{(0)}-K_3^e)^{\prime}+K_1^e \Omega_2^{(0)}-
K_2^e \Omega_1^{(0)}=0.
\end{eqnarray}
Eqs.\ (5-8) determine the  stationary configurations 
and  their corresponding
internal force $f_i^{(0)}$ and torque $(\Omega_i^{(0)}-K_i^e)$ distributions for a 
Kirchhoff filament with spontaneous curvatures.
The stability of a stationary configuration is governed by the DVEs,
which are the lengthy equations  listed in  Appendix A, Eqs.\ (24-29).


The  SSEs and DVEs derived above can be applied in studying the
dynamical properties of any kind of Kirchhoff filaments, however in this paper
we will only study a very special case, the  elastic energy Eq.\ (1),
with $K_1^e=\kappa_{0}$ and $K_2^e=K_3^e=0$. In this case, the SSEs 
Eqs.\  (5-8)
demonstrate that the planar ring configuration is a stationary solution, with
\begin{eqnarray}
&\;&\Omega_1^{(0)}=\kappa={\rm const},
\;\;\;\;\Omega_2^{(0)}=\Omega_3^{(0)}=0,\nonumber \\
&\;&f_1^{(0)}=f_2^{(0)}=f_3^{(0)}=0,
\end{eqnarray}
where $\kappa$ is the (actual) curvature of the ring. 
We now investigate on the dynamical
stability of this ring shape. With Eq.\ (9), the DVEs Eqs.\  (24-29) reduce to
\begin{eqnarray}
&\;&(f_1^{(1)})^{\prime\prime}=\ddot{\alpha}_2, \\
&\;&(f_2^{(1)})^{\prime\prime}-2\kappa (f_3^{(1)})^{\prime}-\kappa^2 f_2^{(1)}
=-\ddot{\alpha}_1,\\
&\;&(f_3^{(1)})^{\prime\prime}-\kappa^2 f_3^{(1)}+2\kappa 
(f_2^{(1)})^{\prime}=0,\\
&\;&\alpha_1^{\prime\prime}-f_2^{(1)}=\ddot{\alpha}_1, \\
&\;&\Gamma \alpha_3^{\prime\prime}+\Gamma\kappa \alpha_2^{\prime}
+\kappa_{0} \alpha_2^{\prime}
-\kappa \kappa_{0} \alpha_3=2 \ddot{\alpha}_3,\\
&\;&\alpha_2^{\prime\prime}-\Gamma\kappa\alpha_3^{\prime}
+(1-\Gamma)\kappa^2\alpha_2-\kappa_{0}
\alpha_3^{\prime}-\kappa_{0} \kappa\alpha_2+f_1^{(1)}
=\ddot{\alpha}_2.
\end{eqnarray}
Redefining the perturbation parameters as $\beta_1=f_1^{(1)}, \beta_2=\alpha_2, 
\beta_3=\alpha_3,  \beta_4=f_2^{(1)}, \beta_5=f_3^{(1)}, \beta_6=\alpha_1$, the
 periodic solutions of this perturbation system Eqs.\ (10-15)
  are of the following form \cite{RJ}
$$
\beta_j={\rm e}^{\sigma t} (x_j {\rm e}^{i n \kappa s}
+x_j^{*}{\rm e}^{-i n \kappa s})\;\;\;(j=1,\cdots,6).
$$
Inserting this into Eqs.\ (10-15),
 we get, in matrix form,
\begin{equation}
{\widehat{\bf{L}}}{\bf \cdot} {\bf x}={\bf 0},
\end{equation}
where ${\bf x}=(x_1, x_2, x_3, x_4, x_5, x_6)^{T}$ and
 ${\widehat{\bf L}}$ is defined as
$$
{\widehat{\bf L}}=\left(
\begin{array}{lr}
{\widehat{\bf L}_1}&{\widehat{\bf 0}}\\
{\widehat{\bf 0}}&{\widehat{\bf L}_2}
\end{array}
\right)
$$
with ${\widehat{\bf 0}}$ being the 3$\times$3 zero-matrix and
\begin{eqnarray}
&\;&{\widehat{\bf L}_1}=\left(
\begin{array}{lcr}
 -n^2 \kappa^2 & -\sigma^2 & 0\\
       0 & i n(\Gamma \kappa+\kappa_{0})\kappa & -\Gamma n^2 \kappa^2 
       -\kappa_{0} \kappa -2\sigma^2\\
       1 & (-n^2 +1-\Gamma) \kappa^2-\kappa_{0} \kappa-\sigma^2 & -i n 
       (\Gamma \kappa+
             \kappa_{0}) \kappa
\end{array}
\right)\\
&\;&{\widehat{\bf L}_2}=\left(
\begin{array}{lcr}
      -1 & 0 & -n^2 \kappa^2-\sigma^2  \\
       2 i n \kappa^2 & -(n^2+1)\kappa^2 & 0 \\
       -(n^2+1)\kappa^2 & -2 i n \kappa^2 & \sigma^2 
\end{array}
\right)
\end{eqnarray}

Eq.\ (16) has non-zero solution ${\bf x}$ if and only
 if the determinant of the matrix $\widehat L$
equals to zero, i.e.,
\begin{equation}
\Delta_L=\Delta_{L_1}\cdot \Delta_{L_2} =0,
\end{equation}
where
\begin{eqnarray}
 \Delta_{L_1}&=&-n^2 (n^2-1)\kappa^4 (\kappa_{0}^2-
 (1-\Gamma)\kappa_{0} \kappa-\Gamma n^2 \kappa^2)
+\sigma^2 ((2+\Gamma) n^4 \kappa^4 \nonumber \\
&\;&-2 (1-\Gamma) n^2 \kappa^4+3 n^2 \kappa_{0} \kappa^3+
\Gamma n^2 \kappa^2 +\kappa_{0} \kappa+
\sigma^2 (2+4 n^2 \kappa^2)),\\
\Delta_{L_2}&=&n^2 (n^2-1)^2 \kappa^6+ \kappa^2 \sigma^2 (n^2+1)+\kappa^4
\sigma^2 (n^2-1)^2.
\end{eqnarray}
$\Delta_{L_2}$ is related the excitation of $\{ f_2^{(1)}, f_3^{(1)}, \alpha_1\}$
as can be inferred from Eq.\ (16) and the characteristics
 of the matrix ${\widehat{\bf L}}$,  
and its value is not related to $\kappa_{0}$,  $\Delta_{L_2} >0$
for any $n\geq 2$. Then to satisfy Eq.\ (19)  we need only
 to consider $\Delta_{L_1}$ which is related to the
excitation of $\{f_1^{(1)}, \alpha_2, \alpha_3\}$. (We note that $\Delta_{L_1}
 (n=1)=\Delta_{L_2}(n=1)=0$ only for $\sigma=0$, therefore the $n=1$ 
modes are just soft modes \cite{RP} not important for
the linear instability.)  

Solution of $\Delta_{L_1}(\sigma)=0$  with real positive $\sigma$ identify the unstable 
modes of $\{f_1^{(1)},\alpha_2, \alpha_3\}$ (these modes
 will be unstable because the amplitudes of the small fluctuations 
will grow exponentially with time.)
 From 
Eq.\ (20) it is evident that
this condition is equivalent to require
\begin{equation}
g=\kappa_{0}^2-(1-\Gamma)\kappa_{0}\kappa-\Gamma n^2\kappa^2\geq 0.
\end{equation}
The general behavior of $g$ is shown in Fig.\ 1. 
It clearly demonstrates that only when the actual
curvature $\kappa$ of a ring is smaller than its spontaneous curvature $\kappa_{0}$
will it be possible for the ring to deform. This prediction is in agreement
with the experiment of Han {\it et al.}, \cite{RA,RB}
 and the threshold curvature for
the $n$-th mode to buckle is
\begin{equation}
\kappa_c(n)= {{-1+\Gamma+\sqrt{(1-\Gamma)^2+4\Gamma n^2}}\over {2 \Gamma n^2}}
\kappa_{0}\;\;\;\;\;\;\;(n\geq 2).
\end{equation}
For $\kappa_{0}=0$, $g$ is always negative 
and no buckling process will occur. Therefore we
can conclude that (i) $\kappa_{0}\neq 0$ and (ii) $\kappa < \kappa_{0}$ are the necessary conditions for the 
instability of a ring  (Fig.\ 1).
 Thus, our present work can qualitatively explain the novel
  phenomenon of DNA described
in the introduction part: $126$-bp DNA will not kink because no matter what
ionic conditions its $\kappa$ is always $\geq\kappa_{c}$; on the other hand,
at some appropriate ionic conditions the curvature of a
 $168$-bp DNA can become  $< \kappa_{c}$, and kink deformation will be
 triggered. \cite{RA,RB}
When taking into account the possibility of nonzero $\omega_0$ in Eq.\ (1), a threshold
condition similar to Eq.\ (23) is derived in Appendix B.

At the initial stage of buckling, only $\{f_1^{(1)}, \alpha_2, \alpha_3\}$ will be excited,
therefore the DNA configuration will deform out of the ring plane rather than deform in 
the same plane as the ring lies in.  Thus in the present we can not tell whether 
the buckled shapes will actually evolve to  the kinked ones observed in experiment.
\cite{RA,RB}  To know this, nonlinear analysis beyond the 
buckling point is needed and it is very much
involved to perform. However, we believe that the linear analysis we employed here has
correctly described the behaviors of DNA at the onset of kinking.

Experiments \cite{RA,RB,RC} shows that Zn$^{2+}$ ions can enhance the instability of DNA 
rings,  while Mg$^{2+}$ ions gives a negative effect. Comparing this with the theoretical
analysis, it is reasonable for us to suggest that Zn$^{2+}$ and 
Mg$^{2+}$ ions  will respectively increase and decrease  the spontaneous
curvature of the DNA chain.  We are informed \cite{RB,RC}
 that that Zn$^{2+}$ mainly binds to
DNA base pairs, and Mg$^{2+}$ interacts with the back-bone phosphate ions. It might be
possible that the intercalation of Zn$^{2+}$ takes place mainly at the exposed side of
the DNA ring  hence causing an increase in its spontaneous curvature. The following
analysis shows  that the
effect of Zn$^{2+}$ to the spontaneous curvature is very significant.

The largest solution of $\Delta_{L_1}(\sigma)=0$ changes with 
 $n$ at four fixed
 $\kappa/\kappa_{0}$ ratios  are shown in Fig.\ 2. The possible deformation
 modes are obviously determined from these curves as they correspond to 
 real positive values of $\sigma$.
  The value of $n$ corresponding to the peaks of these curves ($n_c$)
  are just the most observable modes,\cite{RJ} 
  because these modes grow the fastest. 
 At $1$mM ZnBr$_2$ it is shown experimentally
that $n_c\simeq 4$, \cite{RB}
 therefore we estimate from the solid line of Fig.\ 2
 that $\kappa/\kappa_{0}\simeq 0.25$. The
  value of $\kappa_{0}$ in Zn$^{2+}$-free solutions is till not precisely
 known, but it must be higher than $2\pi/126$ bp$^{-1}$. \cite{RB,RC}
 On the other hand, in this condition $168$-bp is not kinked, so we 
 must have $\kappa\geq \kappa_c(2)$, or in other words,
  $\kappa_{0}\leq 2\pi/90$ bp$^{-1}$ if we set $\Gamma=2/3$ in Eq.\ (23).
Therefore, when there is no Zn$^{2+}$ ions in the solution, $2\pi/126$ bp$^{-1}$
$\leq \kappa_{0}\leq 2\pi/90$ bp$^{-1}$, i.e., $0.55\leq \kappa/\kappa_{0}\leq
0.75$, (the threshold case $\kappa/\kappa_{0}=0.55$ is 
shown in the dot-dashed curve of Fig.\ 2.) 
 Compare this estimate with the value $0.25$
corresponding to $1$mM ZnBr$_2$ we can conclude that
 the addition of $1$mM Zn$^{2+}$ 
ions makes the value of the spontaneous 
 curvature of DNA increase at least one times.

This high efficiency of Zn$^{2+}$  make us  believe it to be possible
that Zn$^{2+}$ ions will induce  spontaneous curvatures to DNAs 
even  when they are originally  linear and 
cause them to kink if they are loaded with stresses. This may be of
vital biological significance. For example, it is known that
transcription and replication of DNA occur with the participation
of specific enzymes which contain the Zn$^{2+}$ ion at the active site.
And it is also found that transcription proceeds simultaneously with
conformational changes of the DNA chain. \cite{VL}  More investigations
on this respect are deserved.

One of the author (Z. H.) appreciates valuable discussions with Dr.\ Yan Jie, Dr.\
Liu Quanhui and Dr.\ Zhao Wei.

\newpage
 
\begin{center}
\Large{Appendix A: The dynamical variation equations (DVEs)}
\end{center}

The dynamical stability of a certain stationary configuration
of the general Kirchhoff filament are determined by the
dynamical variation equations listed below in Eqs.\ (24-29), \cite{RJ,GT}
where $\ddot{(\;)}$ means second order $t$-derivative:
 \begin{eqnarray}
\ddot{\alpha}_2&=&[(\alpha_2^{\prime}-\alpha_3\Omega_1^{(0)}+
\alpha_1\Omega_3^{(0)})^{\prime}-\Omega_1^{(0)}
(\alpha_3^{\prime}-\alpha_1\Omega_2^{(0)}+
\alpha_2\Omega_1^{(0)}) \nonumber \\
&\;&+\Omega_3^{(0)}
(\alpha_1^{\prime}-\alpha_2\Omega_3^{(0)}+
\alpha_3\Omega_2^{(0)})] f_3^{(0)} \nonumber \\
&-&[(\alpha_3^{\prime}-\alpha_1\Omega_2^{(0)}+
\alpha_2\Omega_1^{(0)})^{\prime}-\Omega_2^{(0)}
(\alpha_1^{\prime}-\alpha_2\Omega_3^{(0)}+
\alpha_3\Omega_2^{(0)}) \nonumber \\
&\;&+\Omega_1^{(0)}
(\alpha_2^{\prime}-\alpha_3\Omega_1^{(0)}+
\alpha_1\Omega_3^{(0)})] f_2^{(0)}\nonumber \\
&+&2(\alpha_2^{\prime}-\alpha_3\Omega_1^{(0)}+
\alpha_1\Omega_3^{(0)})({f_3^{(0)}}^{\prime}-f_1^{(0)}\Omega_2^{(0)}+
f_2^{(0)}\Omega_1^{(0)}) \nonumber \\
&-&2 (\alpha_3^{\prime}-\alpha_1 \Omega_2^{(0)}+
\alpha_2\Omega_1^{(0)})({f_2^{(0)}}^{\prime}-f_3^{(0)}\Omega_1^{(0)}+
f_1^{(0)}\Omega_3^{(0)}) \nonumber \\
&+&{f_1^{(1)}}^{\prime\prime}+2 {f_3^{(1)}}^{\prime}\Omega_2^{(0)}-
2 {f_2^{(1)}}^{\prime}\Omega_3^{(0)}
-f_1^{(1)} ((\Omega_3^{(0)})^2+(\Omega_2^{(0)})^2) \nonumber \\
&+&f_2^{(1)}(-
{\Omega_3^{(0)}}^{\prime}+\Omega_2^{(0)}\Omega_1^{(0)})+
f_3^{(1)} ({\Omega_2^{(0)}}^{\prime}+\Omega_3^{(0)}\Omega_1^{(0)}),
\end{eqnarray}
\begin{eqnarray}
-\ddot{\alpha}_1&=&[(\alpha_3^{\prime}-\alpha_1\Omega_2^{(0)}+
\alpha_2\Omega_1^{(0)})^{\prime}-\Omega_2^{(0)}
(\alpha_1^{\prime}-\alpha_2\Omega_3^{(0)}+
\alpha_3\Omega_2^{(0)}) \nonumber \\
&\;&+\Omega_1^{(0)}
(\alpha_2^{\prime}-\alpha_3\Omega_1^{(0)}+
\alpha_1\Omega_3^{(0)})] f_1^{(0)} \nonumber \\
&-&[(\alpha_1^{\prime}-\alpha_2\Omega_3^{(0)}+
\alpha_3\Omega_2^{(0)})^{\prime}-\Omega_3^{(0)}
(\alpha_2^{\prime}-\alpha_3\Omega_1^{(0)}+
\alpha_1\Omega_3^{(0)}) \nonumber \\
&\;&+\Omega_2^{(0)}
(\alpha_3^{\prime}-\alpha_1\Omega_2^{(0)}+
\alpha_2\Omega_1^{(0)})] f_3^{(0)}\nonumber \\
&+&2(\alpha_3^{\prime}-\alpha_1\Omega_2^{(0)}+
\alpha_2\Omega_1^{(0)})({f_1^{(0)}}^{\prime}-f_2^{(0)}\Omega_3^{(0)}+
f_3^{(0)}\Omega_2^{(0)}) \nonumber \\
&-&2 (\alpha_1^{\prime}-\alpha_2\Omega_3^{(0)}+
\alpha_3\Omega_2^{(0)})({f_3^{(0)}}^{\prime}-f_1^{(0)}\Omega_2^{(0)}+
f_2^{(0)}\Omega_1^{(0)}) \nonumber \\
&+&{f_2^{(1)}}^{\prime\prime}-2 {f_3^{(1)}}^{\prime}\Omega_1^{(0)}+
2 {f_1^{(1)}}^{\prime}\Omega_3^{(0)}
-f_2^{(1)} ((\Omega_1^{(0)})^2+(\Omega_3^{(0)})^2) \nonumber \\
&+&f_3^{(1)}(-
{\Omega_1^{(0)}}^{\prime}+\Omega_2^{(0)}\Omega_3^{(0)})+
f_1^{(1)} ({\Omega_3^{(0)}}^{\prime}+\Omega_1^{(0)}\Omega_2^{(0)}),
\end{eqnarray}
\begin{eqnarray}
0&=&[(\alpha_1^{\prime}-\alpha_2\Omega_3^{(0)}+
\alpha_3\Omega_2^{(0)})^{\prime}-\Omega_3^{(0)}
(\alpha_2^{\prime}-\alpha_3\Omega_1^{(0)}+
\alpha_1\Omega_3^{(0)}) \nonumber \\
&\;&+\Omega_2^{(0)}
(\alpha_3^{\prime}-\alpha_1\Omega_2^{(0)}+
\alpha_2\Omega_1^{(0)})] f_2^{(0)} \nonumber \\
&-&[(\alpha_2^{\prime}-\alpha_3\Omega_1^{(0)}+
\alpha_1\Omega_3^{(0)})^{\prime}-\Omega_1^{(0)}
(\alpha_3^{\prime}-\alpha_1\Omega_2^{(0)}+
\alpha_2\Omega_1^{(0)}) \nonumber \\
&\;&+\Omega_3^{(0)}
(\alpha_1^{\prime}-\alpha_2\Omega_3^{(0)}+
\alpha_3\Omega_2^{(0)})] f_1^{(0)}\nonumber \\
&+&2(\alpha_1^{\prime}-\alpha_2\Omega_3^{(0)}+
\alpha_3\Omega_2^{(0)})({f_2^{(0)}}^{\prime}-f_3^{(0)}\Omega_1^{(0)}+
f_1^{(0)}\Omega_3^{(0)}) \nonumber \\
&-&2 (\alpha_2^{\prime}-\alpha_3\Omega_1^{(0)}+
\alpha_1\Omega_3^{(0)})({f_1^{(0)}}^{\prime}-f_2^{(0)}\Omega_3^{(0)}+
f_3^{(0)}\Omega_2^{(0)}) \nonumber \\
&+&{f_3^{(1)}}^{\prime\prime}-2 {f_1^{(1)}}^{\prime}\Omega_2^{(0)}+
2 {f_2^{(1)}}^{\prime}\Omega_1^{(0)}
-f_3^{(1)} ((\Omega_1^{(0)})^2+(\Omega_2^{(0)})^2) \nonumber \\
&+&f_1^{(1)}(-
{\Omega_2^{(0)}}^{\prime}+\Omega_3^{(0)}\Omega_1^{(0)})+
f_2^{(1)} ({\Omega_1^{(0)}}^{\prime}+\Omega_2^{(0)}\Omega_3^{(0)}),
\end{eqnarray}
\begin{eqnarray}
\ddot{\alpha}_1&=&(\alpha_1^{\prime}-\alpha_2\Omega_3^{(0)}+
\alpha_3\Omega_2^{(0)})^{\prime}+(\alpha_2^{\prime}-\alpha_3\Omega_1^{(0)}+
\alpha_1\Omega_3^{(0)})(\Gamma(\Omega_3^{(0)}-K_3^e)-\Omega_3^{(0)})\nonumber\\
&+&K_2^e(\alpha_3^{\prime}-\alpha_1\Omega_2^{(0)}+
\alpha_2\Omega_1^{(0)})+\alpha_2 (\Gamma(\Omega_3^{(0)}-K_3^e)^{\prime}+
K_1^e\Omega_2^{(0)}-K_2^e\Omega_1^{(0)})\nonumber\\
&-&\alpha_3((\Omega_2^{(0)}-K_2^e)^{\prime}-\Gamma(\Omega_3^{(0)}-
K_3^e)\Omega_1^{(0)}+(\Omega_1^{(0)}-K_1^e)\Omega_3^{(0)})-f_2^{(1)}-
\alpha_3 f_1^{(0)},
\end{eqnarray}
\begin{eqnarray}
\ddot{\alpha}_2&=&(\alpha_2^{\prime}-\alpha_3\Omega_1^{(0)}+
\alpha_1\Omega_3^{(0)})^{\prime}+(\alpha_3^{\prime}-\alpha_1\Omega_2^{(0)}+
\alpha_2\Omega_1^{(0)})((\Omega_1^{(0)}-K_1^e)-\Gamma\Omega_1^{(0)})\nonumber\\
&+&(\Omega_3^{(0)}-\Gamma(\Omega_3^{(0)}-K_3^e))(\alpha_1^{\prime}
-\alpha_2\Omega_3^{(0)}+
\alpha_3\Omega_2^{(0)})+\alpha_3 ((\Omega_1^{(0)}-K_1^e)^{\prime}-
(\Omega_2^{(0)}-K_2^e)\Omega_3^{(0)} \nonumber \\
&+&\Gamma (\Omega_3^{(0)}-K_3^e)
\Omega_2^{(0)}))
-\alpha_1(\Gamma(\Omega_3^{(0)}-K_3^e)^{\prime}+K_1^e \Omega_2^{(0)}-
K_2^e \Omega_1^{(0)})+f_1^{(1)}-
\alpha_3 f_2^{(0)},
\end{eqnarray}
\begin{eqnarray}
2 \ddot{\alpha}_3&=&
\Gamma(\alpha_3^{\prime}-\alpha_1\Omega_2^{(0)}+\alpha_2\Omega_1^{(0)})^{\prime}
-K_2^e(\alpha_1^{\prime}-\alpha_2 \Omega_3^{(0)}
+\alpha_3\Omega_2^{(0)})\nonumber\\
&+&K_1^e (\alpha_2^{\prime}-\alpha_3 \Omega_1^{(0)}+\alpha_1\Omega_3^{(0)})
+\alpha_1 ((\Omega_2^{(0)}-K_2^e)^{\prime}-\Gamma(\Omega_3^{(0)}-K_3^e)
\Omega_1^{(0)}+(\Omega_1^{(0)}-K_1^e)\Omega_3^{(0)})\nonumber\\
&-&\alpha_2 ((\Omega_1^{(0)}-K_1^e)^{\prime}-(\Omega_2^{(0)}-K_2^e)\Omega_3^{(0)}
+\Gamma(\Omega_3^{(0)}-K_3^e)\Omega_2^{(0)})+\alpha_1 f_1^{(0)}+\alpha_2
f_2^{(0)}.
\end{eqnarray}

\newpage
\begin{center}
\Large{Appendix B: The linear instability result for Eq.\ (1) with nonzero $\omega_0$}
\end{center} 

In the main text, we have focused our attention to the $\omega_0 =0$ case  of medel Eq.\ (1). 
For the general case of nonzero $\omega_{0}$, following the
same procedure as discussed in the main text  we can obtain the threshold condition for the $n$-th mode
to become instable. It reads
\begin{eqnarray}
&\kappa^9(n^2-1) n^4 (n+1)^2\times \nonumber \\
 &\left(\kappa^2\kappa_{0}(\Gamma-1)(1-n^2)+\kappa\kappa_{0}^2(1-n^2)
-\Gamma^2\omega_0^2(\kappa_{0}+\Gamma\kappa n^2)+\Gamma\kappa^3 n^2(n^2-1)\right)=0.
\end{eqnarray}
In the limiting case of $\omega_0=0$ (the filament has intrinsic bend but no intrinsic twist),
 Eq.\ (30) reduces to Eq.\ (23) of the main text. 
In the limiting case of $\kappa_{0}=0$ (then DNA has intrinsic twist but no intrinsic bend),
Eq.\ (30) reduces to
\begin{equation}
-\Gamma\kappa^{10}n^6(n^2-1)^2 \left( \Gamma^2 \omega_0^2-\kappa^2(n^2-1)\right).
\end{equation}
Eq.\ (31) is just the same result obtained in previous references for a chain with intrinsic twist,\cite{R13,R14,R15}
showing the correctness of this method.


\begin{figure}
\caption{The  behavior of $g$ [Eq.\ (22)] for $n=2$ and $\Gamma=2/3$.
 $g$ is
negative for $\kappa_{0}/\kappa\leq 1$ (the dotted part of the line), 
it becomes positive only when $\kappa_{0}/\kappa$ exceeds some 
threshold value higher than
$1$ (the solid part).  $g(\kappa_{0}/\kappa=1)=-\Gamma (n^2-1)$  and is 
negative for the important case of $n
\geq 2$. Thus it is obvious that
buckling process can take place only for $\kappa <\kappa_{0}$. 
This prediction is 
confirmed by experiment.$^{1,2}$ }
\label{FIG. 1}
\end{figure}

\begin{figure}
\caption{
The relation between the largest solution of $\Delta_{L_1} (\sigma)=0$ 
[Eq.~(20)]
and $n$, at  $\kappa/\kappa_{0}=0.25$ (the solid line), $0.30$
 (the dotted line), 
$0.40$ (the dashed line), and $0.55$ (the dot-dashed line)
 for the $168$-bp DNA ring. 
We set $\kappa=\pi/(0.34* 168)$ for the $168$-bp DNA$^{10,12-15}$
 and choose $\Gamma=2/3$.$^{10,15}$
The most dominant deformation mode determined by the peaks
 of these curves increases 
with the decreasing  of  $\kappa/\kappa_{0}$, and the number 
of observable modes also
increases as $\kappa/\kappa_{0}$ decreases. 
At $\kappa/\kappa_{0}=0.25$, the dominant mode is
$n=4$ and the modes $n=2, 3, 4$ can be observed; while at $\kappa/\kappa_{0}
=0.55$, no deformation mode with $n\geq 2$ will be excited. 
This prediction is in very close agreement
with experimental observations.$^2$
}
\label{FIG. 2} 
\end{figure}

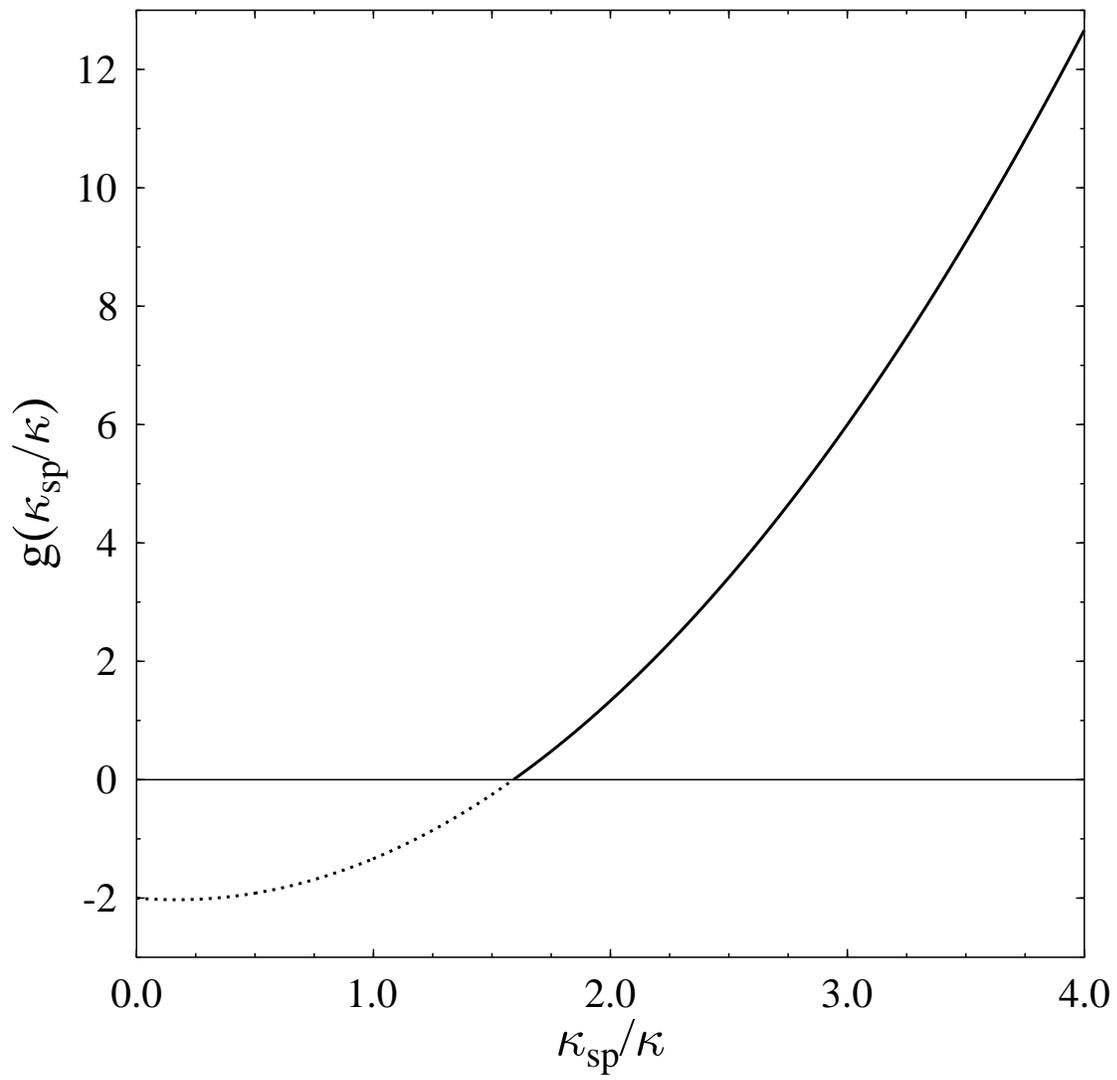
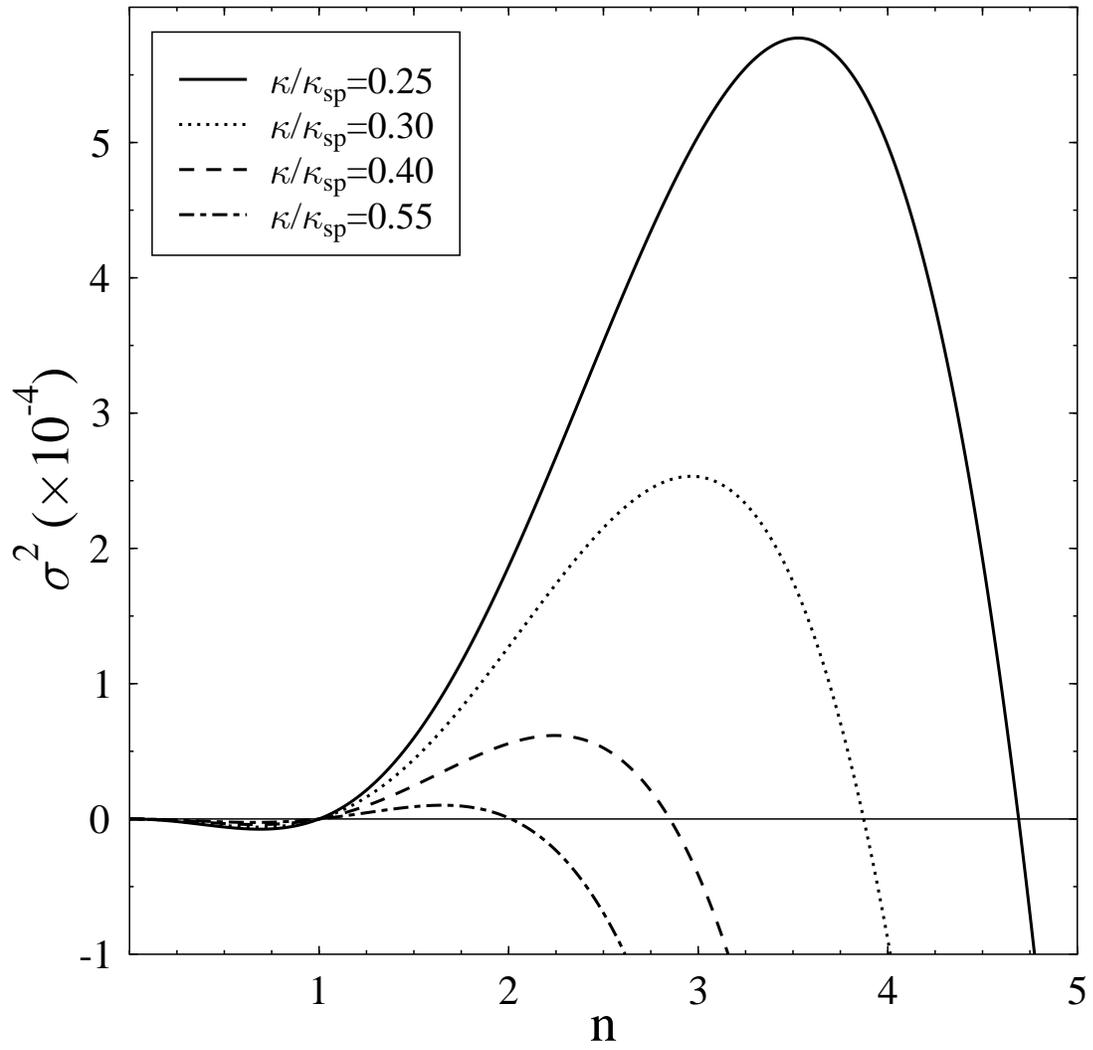
\end{document}